\documentclass[runningheads,orivec]{llncs}
\usepackage[T1]{fontenc}
\usepackage{graphicx,verbatim}
\usepackage[acronym]{glossaries}
\usepackage{orcidlink}
\usepackage{multirow}
\usepackage[misc,geometry]{ifsym}

\newacronym{iac}{IAC}{intracranial arterial calcification}

\begin{document}
\title{Automated Distinction of Intimal and Medial Intracranial Arterial Calcification from CT Head}
\titlerunning{Automated Distinction of Intracranial Arterial Calcification Subtype}
\author{Benjamin Jin\inst{1,2}\textsuperscript{(\Letter)}\orcidlink{0009-0006-2710-7248} \and
Maria del C. Valdés Hernández\inst{1,3}\orcidlink{0000-0003-2771-6546} \and
Richard Bortsov\inst{4}\orcidlink{0000-0002-7614-8437} \and
Joanna M. Wardlaw\inst{1,3}\orcidlink{0000-0002-9812-6642} \and 
Daniel Bos\inst{2,4}\orcidlink{0000-0001-8979-2603} \and
Grant Mair\inst{1,5}\orcidlink{0000-0003-2189-443X}
}
\authorrunning{B. Jin et al.}
\institute{
    Institute for Neuroscience and Cardiovascular Research, University of Edinburgh, United Kingdom\\
        \email{b.jin@ed.ac.uk}
    \and Department of Epidemiology, Erasmus MC, The Netherlands
    \and UK Dementia Research Institute, University of Edinburgh, United Kingdom
    \and Department of Radiology and Nuclear Medicine, Erasmus MC, The Netherlands
    \and Department of Clinical Neurosciences, NHS Lothian, United Kingdom
}

\maketitle
\begin{abstract}
\Acrfullpl{iac} are a common finding on clinical non-contrast enhanced head CT scans and are associated with neurovascular disease. Calcifications can occur in the intimal or medial layer of the arterial wall, subtypes that differ in aetiology and may have distinct clinical relevance. These subtypes can be visually distinguished by radiologists based on the shape of the calcifications. We investigate three automated approaches for subtype classification of \acrshort{iac} from head CT-derived segmentation masks: (1) an automated adaptation of the established radiological visual score, (2) a sphericity-based method, and (3) a method based on shape embeddings extracted by a medical shape foundation model. All approaches use the same lightweight classification pipeline on top of the features they compute and are evaluated using 5-fold cross-validation. The three methods achieved comparable performance, with the embedding-based approach yielding the best overall results with a weighted F1 (mean $\pm$ SD) of up to $ 71.5 \pm 3.7 $ for a single artery and 59.8 $\pm$ 1.7 for the joint artery classification. Performance was largely preserved when using automated instead of manual \acrshort{iac} segmentation masks, and we found the difference in weighted F1 not significant. Our results show that fully automated \acrshort{iac} subtype quantification from head CT is feasible and remains robust to the use of manual and automated \acrshort{iac} segmentation masks. Code at \url{https://github.com/bjin96/iac-subtyping}


\end{abstract}

\section{Introduction}
\begin{figure}[t!]
\includegraphics[width=\textwidth]{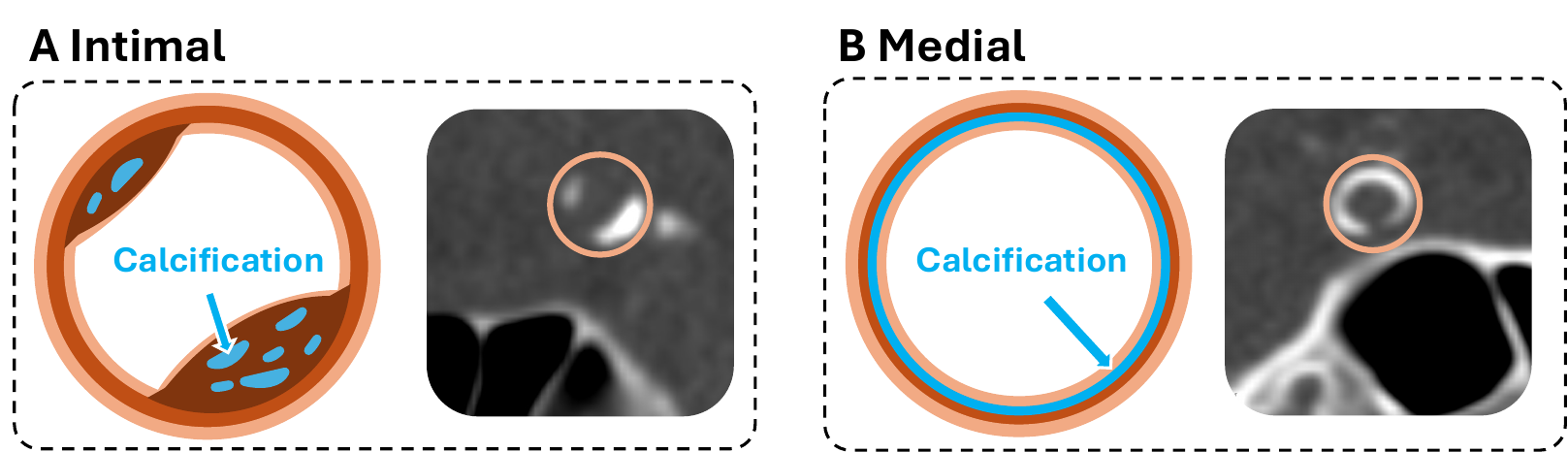}
\caption{Intracranial arterial calcification subtypes with examples of the calcification from the cavernous part of the internal carotid artery as it appears on CT.} \label{figure:intimal_medial_calcification}
\end{figure}

\Acrfull{iac} is visible on routine clinical head CT scans and associated with neurovascular diseases (e.g. stroke \cite{Banerjee17}, dementia \cite{Chen19}). These associations have primarily been based on \acrshort{iac} presence and volume, but a growing body of evidence suggests that different calcification shapes reflect distinct processes with divergent downstream consequences \cite{Du22,Bartstra20}. The main \acrshort{iac} subtypes are intimal and medial \acrshort{iac} \autoref{figure:intimal_medial_calcification}. Intimal calcification occurs as part of the atherosclerotic process, in which the calcium accumulates alongside fatty and fibrous materials within the intimal layer of the arterial wall \cite{Libby19}. Medial calcification is often used to describe calcification of the internal elastic lamina and the medial layer of the arterial wall. Medial calcification is thought to lead to arterial stiffening with an influence on haemodynamics \cite{Kockelkoren17}.

Kockelkoren \textit{et al.} previously developed a visual score to distinguish between the two subtypes \cite{Kockelkoren17}. Here, we propose different approaches -- from visual score-inspired methods to using shape embeddings from a shape foundation model -- for the automated distinction of intimal and medial calcification from \acrshort{iac} segmentation masks. These masks can be created manually or via automated segmentation \cite{Bortsova21,Jin26}. To our knowledge, this is the first study to investigate automated classification of intimal and medial \acrshort{iac} from head CT.

\section{Methods}

\begin{figure}[t!]
\includegraphics[width=\textwidth]{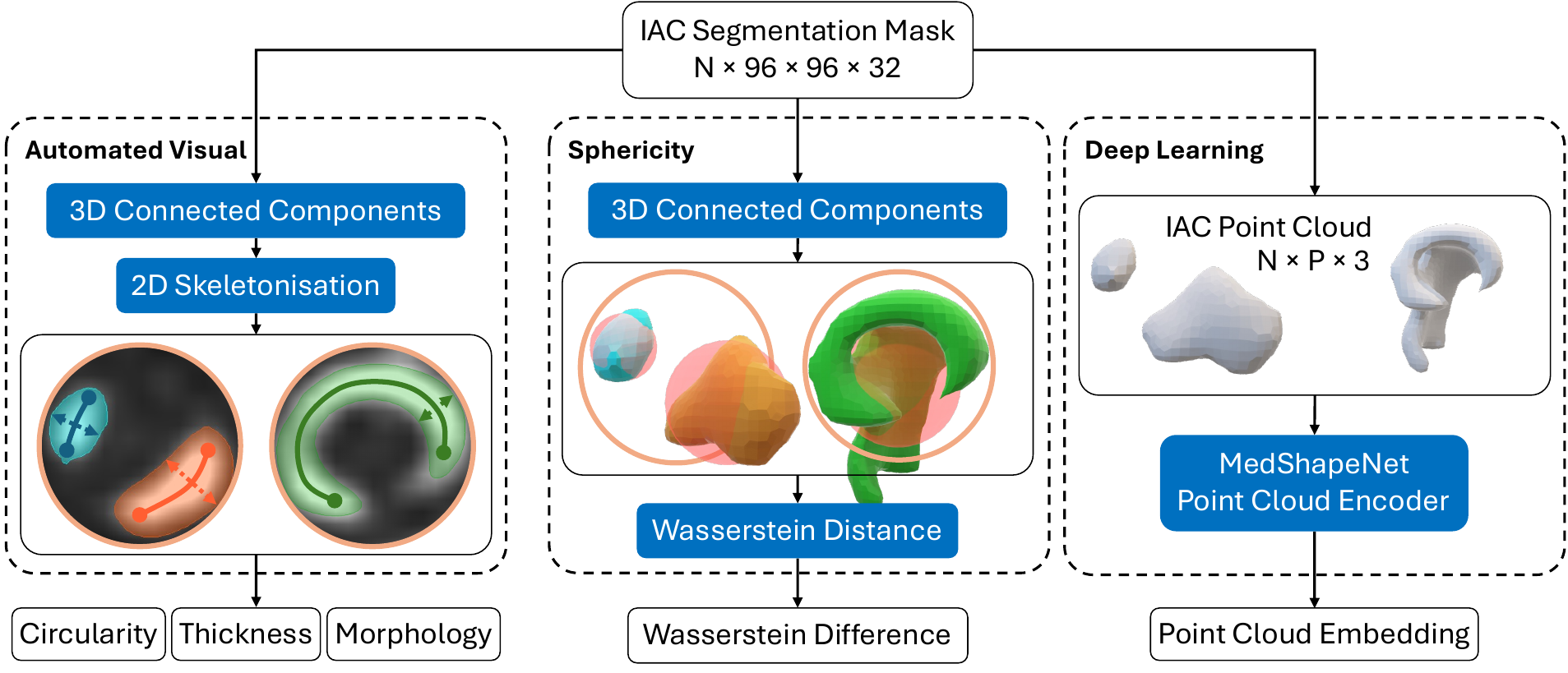}
\caption{Three methods generating features for distinguishing \acrshort{iac} subtypes. Spheres for sphericity calculation overlayed in red over the \acrshort{iac} objects. $N$ is the number of examples, $P$ is the respective number of points derived from the segmentation masks.} \label{figure:methods_overview}
\end{figure}

We describe the dataset and automated methods generating features of \acrshort{iac} subtypes (\autoref{figure:methods_overview}). We assume the availability of \acrshort{iac} segmentation masks. 

\subsection{Data}

The Rotterdam Study \cite{Ikram17} is a prospective, population-based cohort study investigating diseases prevalent in ageing populations.
The Rotterdam Study has been approved by the Medical Ethics Committee of Erasmus MC (registration no. MEC 02.1015) and by the Dutch Ministry of Health, Welfare and Sport (Population Screening Act WBO, license no. 1071272-159521-PG). All participants provided written informed consent. The Rotterdam Study data may be requested through a formal data access procedure. Data access for this work was granted under a data use agreement.
As part of the study, 2524 participants underwent a non-contrast CT scan between 2003 and 2006. \acrshort{iac} was manually segmented \cite{Bos12} and intimal and medial calcification subtypes were visually assessed \cite{Beukel22} following criteria introduced in \cite{Kockelkoren17}. Labels were available separately for the left and right internal carotid artery as binary intimal/medial subtype labels, and jointly at the patient level as a global predominant subtype label with an additional \textit{mixed} category. The mixed category indicates that each subtype was assigned to one of the individual internal carotid arteries.

The automated segmentation masks were generated using preprocessed CT scans from \cite{Bortsova21} and an ensemble of ViTiac-S models from \cite{Jin26} fine-tuned with 5-fold cross-validation on 100 CT image series from The Rotterdam Study. The CT image series used during fine-tuning were excluded from further analysis. 2219 CT image series from the same number of participants were available at the end of the segmentation pipeline. CT scans without \acrshort{iac} cannot be subtyped, so we use only examples that exhibit \acrshort{iac}. We also exclude cases with minimal calcification (less than 4 mm$^3$), as they can lead to instability, mainly when computing shape embeddings. Manual and automated segmentations with \acrshort{iac} present were available from 1766 and 1704 participants, respectively.

\subsection{Automated Visual Method}

As a baseline, we automatically compute features inspired by the visual score described by Kockelkoren \textit{et al.} \cite{Kockelkoren17}. From the binary segmentation masks, we compute 3D connected components, skeletonise the calcification objects on the 2D slices (axial, coronal, and sagittal views), and extract skeleton paths. We then assess three characteristics analogous to the manual visual assessment.
1) \textit{Circularity} as the mean of the maximum sum of same-sign turn angles along each skeleton path through each calcification object. Due to the skeletonisation and smoothing of the skeleton path to counteract noise, the path length shrinks, leading to slightly decreased maximum angles,
2) \textit{Thickness} as the mean thickness of the calcification objects along their longest 2D skeleton path,
and 3) \textit{Morphology} as the mean length of the longest 2D path per calcification object. We use the length of calcification objects as a proxy measure for the continuity assessed in the original visual score.
These features are first computed at the level of calcification objects and then averaged across objects within each CT image series to obtain scan-level features for classification.

\subsection{Sphericity Method}

In addition to the Automated Visual method, we develop a simpler method for describing the \acrshort{iac} subtype that is based solely on the \textit{sphericity} \cite{Wadell1935} of the 3D \acrshort{iac} shapes. Sphericity $\Psi$ measures the ratio between the surface of a sphere with the same volume as the considered 3D object and its actual surface area:

\begin{equation}
    A_s = \pi^{\frac{1}{3}}(6V_{p})^{\frac{2}{3}}, \quad \Psi = \frac{A_s}{A_p}.
\end{equation}

$V_p$ and $A_p$ are the volume and the surface area of a given calcification object $p$. $V_s$ and $A_s$ are the volume and surface area of a sphere $s$ with $V_s = V_p$.

A CT image series has potentially multiple calcification objects. Therefore, the output of the sphericity calculation is a distribution of sphericities across these objects after the connected component analysis. We calculate Wasserstein (or Earth Mover's) distances \cite{Villani2009} between the CT image series and the overall distribution for each \acrshort{iac} subtype across the training folds. The Wasserstein distance quantifies the cost of transforming one distribution into another. Based on it, we define a Wasserstein difference $w_\Delta \in [-1, 1]$ as:

\begin{equation}
    w_{\Delta,c} = \mathrm{wasserstein}(D_c, D_i) - \mathrm{wasserstein}(D_c, D_m).
\end{equation}

$D_c$ is the sphericity distribution within a single CT image series $c$, and $D_i$ and $D_m$ are the reference sphericity distributions of calcification objects from training cases labelled as predominantly intimal and predominantly medial, respectively. We utilise $w_\Delta$ as a feature for the classification pipeline.

\subsection{Deep Learning Method}

In a more exploratory computational approach, we use embeddings from a 3D point cloud foundation model to obtain shape descriptors without designing subtype-specific hand-crafted features. We input point clouds derived from \acrshort{iac} segmentation masks for each arterial side. I.e., the input contains all calcification objects from a single internal carotid artery. As a foundational model, we experiment with MedShapeNet \cite{Yassin24}, which is trained to complete a wide variety of medical shapes \cite{Li25}. Here, we use only the Point Cloud Encoder from MedShapeNet to obtain point cloud embeddings as shape features for classification, discarding other parts of the original model (e.g. the Text Encoder and the Point Cloud Decoder). We reduce the embedding dimension $E$ from the original 4096 to 128.

\subsection{Classification Pipeline}
For all three approaches, we construct the same lightweight classification pipeline and compute common classification metrics to compare the descriptiveness of their features. The pipeline performs logistic regression and 5-fold cross-validation using the popular scikit-learn library \cite{Pedregosa11}. We train and evaluate the pipeline separately for each of the arteries (left and right) and for the joint arteries (total). For total classification, we concatenate automated visual features, sphericity features, and MedShapeNet embeddings from both sides for the respective methods.

\section{Results}

We evaluate the classification performance of our three proposed automated methods for distinguishing intimal and medial \acrshort{iac} subtypes and inspect the impact of using manually or automatically generated \acrshort{iac} segmentation masks.

\subsection{Intimal and Medial Calcification Classification}

\begin{table}[t!]
\setlength{\tabcolsep}{0.3em} 
\renewcommand{\arraystretch}{1.1}
\centering
\begin{tabular}{l|c||r|r|r|r|r|r|r}
    \multicolumn{2}{c||}{} & \multicolumn{3}{c|}{Macro Avg.$\pm$SD} & \multicolumn{4}{c}{Weighted Avg.$\pm$SD} \\
    \cline{3-9}
    Method & Side & \textbf{F1} & \textbf{Prec.} & \textbf{Recall} & \textbf{F1} & \textbf{Prec.} & \textbf{Recall} & $\mathrm{\textbf{F1}}^{\mathrm{\textbf{A}}}$\\
    \hline
    \hline
    \multirow{3}{*}{\shortstack{Autom.\\Visual}}

& L & 66.5$\pm$2.7 & 67.7$\pm$2.3 & 68.8$\pm$2.5 & 67.3$\pm$2.7 & 71.1$\pm$2.2 & 66.8$\pm$2.7 & 65.3$\pm$1.9 \\
& R & 68.4$\pm$2.1 & 69.5$\pm$2.1 & 70.3$\pm$2.2 & 68.9$\pm$2.1 & 72.0$\pm$2.2 & 68.6$\pm$2.1 & 66.2$\pm$1.9 \\
& T & 49.9$\pm$1.8 & 51.3$\pm$1.6 & 51.1$\pm$2.1 & 58.4$\pm$1.8 & 62.7$\pm$1.5 & 56.7$\pm$2.0 & 57.5$\pm$2.5 \\
    \hline
    \multirow{3}{*}{Spher.}
& L & 67.3$\pm$1.2 & 67.5$\pm$1.1 & 68.5$\pm$1.2 & 68.5$\pm$1.3 & 70.4$\pm$1.1 & 68.0$\pm$1.3 & 69.0$\pm$5.1 \\
& R & 67.5$\pm$3.0 & 67.5$\pm$3.0 & 68.1$\pm$3.2 & 68.6$\pm$3.0 & 69.5$\pm$3.0 & 68.3$\pm$3.0 & 69.1$\pm$2.6 \\
& T & 50.3$\pm$3.1 & 52.3$\pm$2.9 & 51.4$\pm$3.9 & 58.7$\pm$2.4 & 63.3$\pm$2.6 & 56.0$\pm$2.6 & 58.0$\pm$2.0 \\
    \hline
    \multirow{3}{*}{\shortstack{Deep\\Learn.}}
& L & 67.8$\pm$0.7 & 67.7$\pm$0.7 & 68.5$\pm$0.9 & \textbf{69.4}$\pm$0.7 & 70.3$\pm$0.8 & 69.1$\pm$0.7 & 69.5$\pm$1.8 \\
& R & 70.2$\pm$3.9 & 70.3$\pm$4.0 & 70.3$\pm$3.8 & \textbf{71.5}$\pm$3.7 & 71.7$\pm$3.7 & 71.5$\pm$3.8 & 71.0$\pm$3.7 \\
& T & 50.8$\pm$1.9 & 51.5$\pm$1.5 & 51.2$\pm$2.1 & \textbf{59.8}$\pm$1.7 & 62.1$\pm$1.3 & 58.0$\pm$2.1 & 57.9$\pm$1.7 \\
\end{tabular}
\caption{\Acrshort{iac} subtype classification cross-validation metrics for the three proposed methods, mainly from manual segmentations. Macro Avg. is the unweighted mean of per-class metrics, and Weighted Avg. is weighted by per-class sample count. We provide separate results for the left (L) and the right (R) artery, and joint total (T) results. Weighted F1 using \textit{automatically} generated segmentations in column $\mathrm{\textbf{F1}}^{\mathrm{\textbf{A}}}$.}
\label{tab:classification_results}
\end{table}

We evaluate \acrshort{iac} subtype classification from manual segmentation masks in \autoref{tab:classification_results}, showing classification performance independent of variability in segmentation quality from different segmentation algorithms. We separately test the classification of calcification subtypes in the left, right, and joint internal carotid arteries against the reference manual visual assessment.

\begin{table}[b!]
\centering
\begin{tabular}{l|l||r|r|r|r|r|r}
    \multicolumn{2}{c||}{} & \multicolumn{2}{c|}{Left} & \multicolumn{2}{c|}{Right} & \multicolumn{2}{c}{Total} \\
    \cline{3-8}
    \multicolumn{2}{c||}{Method Pair} & $\kappa$ & Agree(\%) &  $\kappa$ & Agree(\%) & $\kappa$ & Agree(\%) \\
    \hline
    \hline
        Autom. Visual & Sphericity & \textbf{0.58} & \textbf{79.1\%} &\textbf{ 0.56} & \textbf{77.8\%} & \textbf{0.46} & \textbf{64.3\%} \\
        Autom. Visual & Deep Learning & 0.38 & 68.8\% & 0.42 & 70.7\% & 0.32 & 56.0\% \\
        Sphericity & Deep Learning & 0.39 & 69.6\% & 0.43 & 72.1\% & 0.32 & 55.8\% \\
\end{tabular}
\caption{Cohen's $\kappa$ and raw agreement percentages for the pairwise agreement between the proposed Automated Visual, Sphericity, and Deep Learning methods.}
\label{tab:score_agreement}
\end{table}

\begin{figure}[t!]
\includegraphics[width=\textwidth]{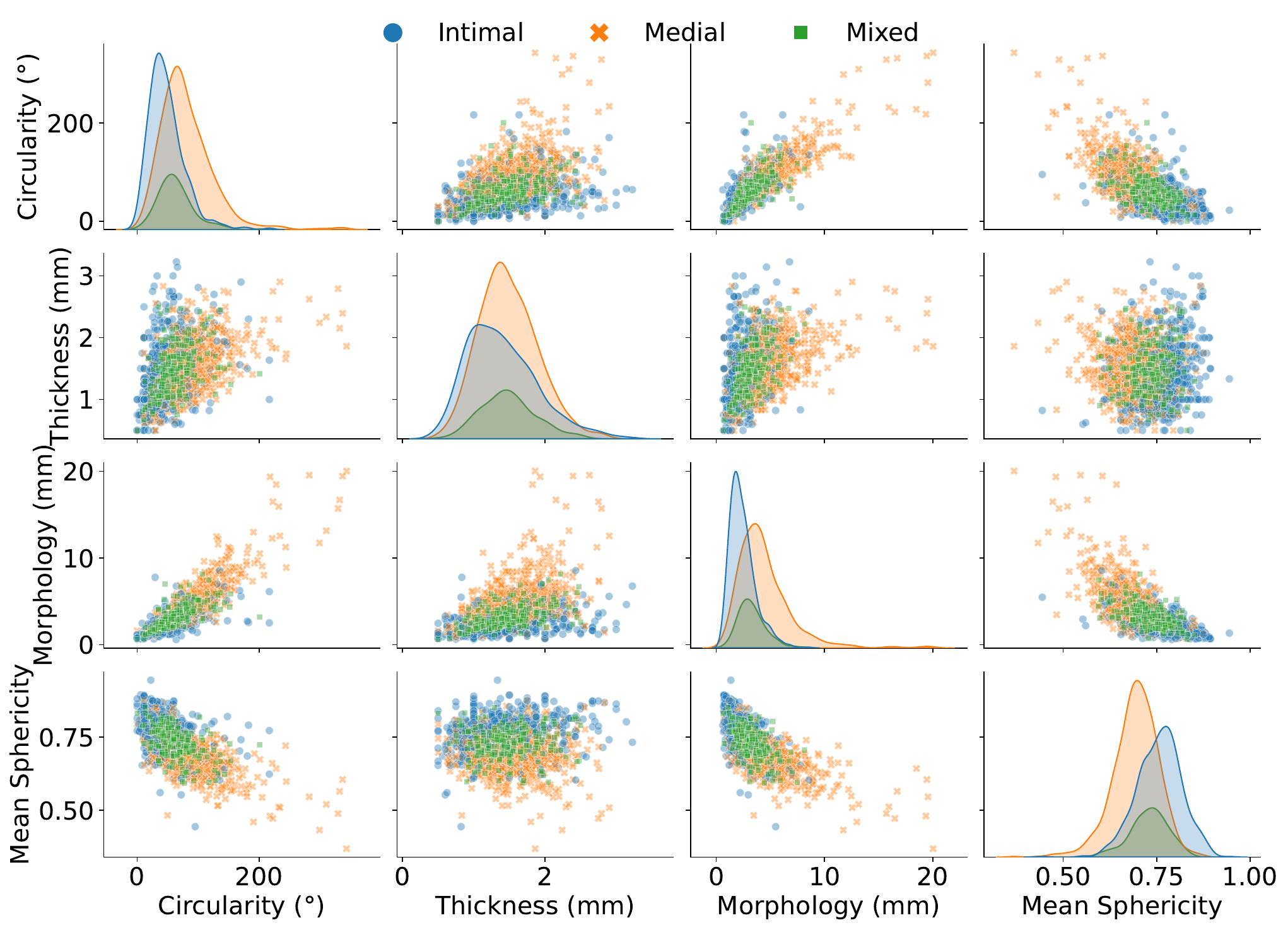}
\caption{Correlation between Automated Visual features and mean sphericity.} \label{figure:kockelkoren_sphericity_correlation}
\end{figure}

While the different methods achieve relatively similar results in distinguishing intimal and medial \acrshort{iac} subtypes, we see differences in their pairwise agreement (\autoref{tab:score_agreement}). The Automated Visual and the Sphericity methods exhibit the largest agreement, whereas both have decreased agreement with the Deep Learning method. Inspecting the correlation between sphericity measures and Automated Visual features, we find that the agreement is likely related to a moderate to strong relationship between the mean sphericity and the computed circularity (Spearman correlation $\rho = -0.62$) and morphology ($\rho = -0.70$). There is no correlation between mean sphericity and thickness ($\rho = 0.00$), but thickness appears less discriminative visually (\autoref{figure:kockelkoren_sphericity_correlation}). The lower agreement between the Deep learning method and the two hand-crafted approaches suggests that it may capture complementary shape characteristics beyond those captured in the Automated Visual and Sphericity features.

\begin{figure}[h!]
\includegraphics[width=\textwidth]{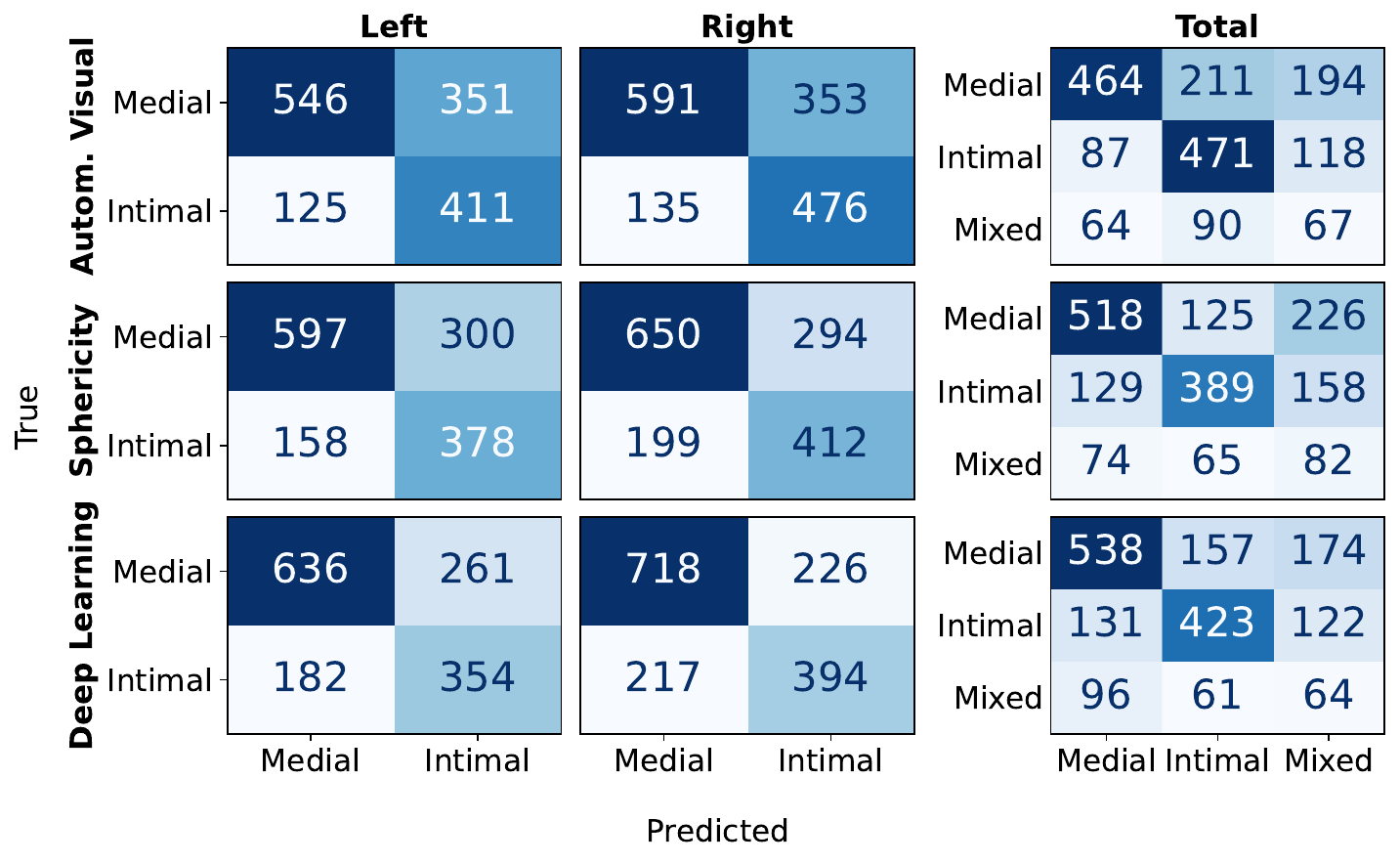}
\caption{Confusion matrices for artery-wise classification of IAC subtype using our Automated Visual, Sphericity, and Deep Learning methods.} \label{figure:manual_confusion_matrices}
\end{figure}

We observe a tendency of the Automated Visual method to confuse medial for intimal \acrshort{iac} (\autoref{figure:manual_confusion_matrices}). The Sphericity and the Deep Learning methods exhibit a better balance between the classification performance of the two subtypes. All three methods struggle to identify the \textit{mixed} class, which denotes different predominant subtypes in individual internal carotid arteries. 

\subsection{Manual and Automated Segmentation Masks}

\begin{figure}[hb!]
\includegraphics[width=\textwidth]{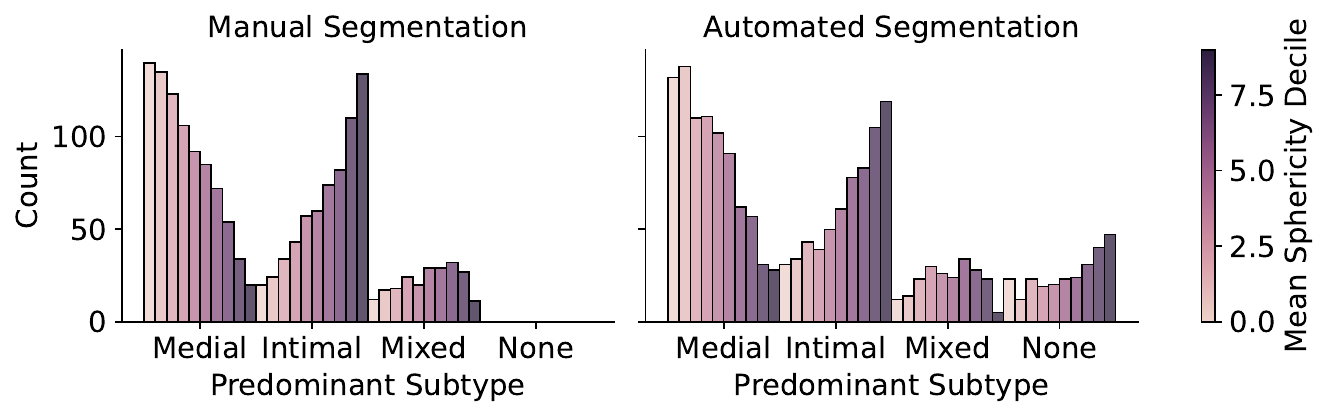}
\caption{Deciles summarising the scan-level mean sphericity for each predominant subtype. The additional None category for automated segmentation masks includes false-positive segmentations and calcifications missed by human annotators.} \label{figure:sphericity_distributions}
\end{figure}

Beyond classification performance based on manual \acrshort{iac} segmentation masks, we evaluate the impact of using automatically generated segmentation masks. As an example of the impact on segmentation-derived features, \autoref{figure:sphericity_distributions} shows the distribution of mean sphericity per CT image series in deciles derived from manual and automated segmentation masks. The intimal and medial distributions show slight differences between the segmentation origins, but keep the same contrasting distribution shapes. To compare classification metrics using manual and automated segmentation masks, we perform bootstrapping (10,000 resamples) on the respective predictions. We calculate the 95\% CI of the mean differences between weighted F1 scores and find that the results are not significant.

\subsection{Discussion and Limitations}
Manual visual assessment is a practical way to distinguish \acrshort{iac} subtypes from head CT scans for small to medium-sized datasets \cite{Kockelkoren17}. Yet, visual scores have several downsides, including a relatively coarse scale, descriptions at the level of arteries rather than individual \acrshort{iac} objects, and linearly increasing manual labour as the size of the study population grows. This makes them (potentially prohibitively) costly for population-level datasets. Related to our work, morphometric \acrshort{iac} quantification methods have recently been published, but rely on CT angiography \cite{Berghout25} and lack information about \acrshort{iac} subtypes \cite{Xu26}. For the first time, we present automated distinction of \acrshort{iac} subtypes.

The Automated Visual method follows the features of the manual visual assessment introduced in \cite{Kockelkoren17} that has been validated against histological findings. However, the calculation is based on the 2D shape of the 3D \acrshort{iac} objects and on correlated features (e.g. circularity and long continuous paths). The Sphericity method simplifies the criteria described by Kockelkoren \textit{et al.} into a single 3D measurement. Computing the Wasserstein distance between the sphericities in individual CT image series to the \acrshort{iac} subtype distribution in the training data, the Sphericity method adapts to empirical data. Yet, it is still bound to the heuristic that intimal calcifications appear more spherical than medial calcifications. The Deep Learning method follows a fully data-driven approach. Rather than injecting specific shape information into the model, the shape foundation model extracts general shape representations from the segmentation mask. In addition to increased flexibility in shape descriptions, this allows for the representation of combinations of calcification objects, albeit at the cost of model explainability. The point cloud encoder could be adapted for tiny objects, or they could be handled separately in the classification pipeline in the future.

Our approach is compatible with integrating automated \acrshort{iac} segmentation to form a fully automated end-to-end pipeline from head CT scan to the quantification and distinction of intimal and medial calcification subtypes. Fully automated \acrshort{iac} description enables large-scale quantification of \acrshort{iac} subtypes and downstream association with neurovascular pathology and clinical outcomes.

\section{Conclusion}
In this work, we proposed three approaches to automating the distinction between intimal and medial \acrfull{iac}. In combination with automated \acrshort{iac} segmentation, our work supports the fully automated quantification of \acrshort{iac} subtypes directly from CT scans. Given the different aetiologies of \acrshort{iac} subtypes and evidence of distinct downstream consequences for neurovascular diseases, their automated quantification has potential for the application on large-scale clinical datasets. Applications could include generating population-level evidence linking \acrshort{iac} and neurovascular diseases, or the targeted recruitment of at-risk participants for research studies into treatments for neurovascular diseases. In the longer term, automated \acrshort{iac} subtype quantification could inform individual patient stratification in clinical practice.

\begin{credits}
\subsubsection{\ackname}
B. J. gratefully acknowledges funding from the SINAPSE ECR Exchange Fund 2025 and the Doctoral Training Programme in Precision Medicine from the Medical Research Council [MR/W006804/1].
M.V.H. is funded by The Row Fogo Charitable Trust [BROD.FID3668413].
J.M.W is part-funded by the UK Dementia Research Institute, which is funded by the UK MRC, Alzheimer's Society and Alzheimer's 
Research UK, and by the UK National Institute of Health and Care Research. 
\subsubsection{\discintname}
The authors have no competing interests to declare that are relevant to the content of this article.
\end{credits}
 
\bibliographystyle{splncs04}
\bibliography{references}

\end{document}